\documentstyle[preprint,revtex,12pt]{aps}

\begin{document}
\begin{title}
\baselineskip30pt
Critical Exponents of the Three Dimensional\\
Random Field Ising Model
\end{title}
\renewcommand{\thefootnote}{\fnsymbol{footnote}}
\addtocounter{footnote}{1}\footnotetext{Present address: Institut f\"ur
Theoretische Physik, Universit\"at zu K\"oln,\\\hskip0.25cm
5000 K\"oln 41, Germany.}
\author{H.\ Rieger\fnsymbol{footnote} and A.\ Peter Young}
\begin{instit}
Physics Department\\
University of California\\
Santa Cruz, CA 95064
\end{instit}
\begin{abstract}
\baselineskip18pt
The phase transition of the three--dimensional random field Ising model
with a discrete ($\pm h$) field distribution is investigated by extensive
Monte Carlo simulations. Values of the critical exponents for the
correlation length, specific heat, susceptibility, disconnected
susceptibility and magnetization are determined simultaneously via
finite size scaling. While the exponents for the magnetization and
disconnected susceptibility are consistent with a first order transition,
the specific heat appears to saturate indicating no latent heat. Sample
to sample fluctuations of the susceptibilty are consistent with the
droplet picture for the transition.
$\quad$ 
\vskip3cm
\begin{center}
Submitted to {\em Europhysics Letters}
\end{center}
\end{abstract}
\narrowtext
\baselineskip22pt
The three dimensional ferromagnetic Ising model with a random field
(RFIM) shows a phase transition to long range order at a critical temperature
for small enough field strength \cite{BY}.
However, the nature of this
transition is still unclear; even the question of whether it is
first \cite{YN} or second \cite{OgHu,Og} order remains unsettled.
The droplet theory of Villain \cite{Vi} and Fisher \cite{Fi}
(see also Bray and Moore \cite{BM})
develops a self--consistent picture of the transition as well as a set
of scaling relations among the critical exponents.
Existing numerical studies have been unable to test the validity
of these scaling relations because not all the
exponents were calculated for any of the relations.
The aim of this paper is to determine all critical exponents within
a single numerical simulation in order to test the scaling relations
predicted by the droplet picture.

The droplet picture makes other predictions
which are relevant to our simulations. One is that at and below the
transition temperature $T_c$, the susceptibility is expected to have
large sample to sample fluctuations \cite{DSY}. We therefore need to
average over a large number of samples to get a good statistics.
Another prediction is thermally activated dynamical scaling \cite{Fi,Vi}
resulting in
a dramatic slowing down in the critical region. This means very long
equilibration times. For these reasons we had to confine
ourself to modest lattice sizes and the critical exponents will
be obtained via finite size scaling.

The Hamiltonian of the system is given by
\begin{equation}
{\cal H}=-\sum_{\langle ij\rangle}S_i S_j - \sum_i h_i S_i\;,
\end{equation}
where $S_i=\pm1$ are Ising spins and the first sum runs over all
nearest neighbor pairs on an $L\times L\times L$ simple cubic lattice
with periodic boundary conditions. The random fields $h_i$ in the
second sum, running over all sites, take random values with the discrete
probability distribution
\begin{equation}
P(h_i) = \frac{1}{2}\delta(h_i-h_r)+\frac{1}{2}\delta(h_i+h_r)\;.
\end{equation}
The Monte Carlo (MC) simulations were performed on a transputer array
using 40 T414 transputers. We were able to obtain high performance by
using a multi-spin coding algorithm described in
\cite{Ri} in which each transputer simulates 32
physically different systems in parallel, each with
a different random field realization. This is somewhat different
from the implementation of multi-spin coding which was applied to the RFIM
in \cite{shab84}.
For each run at fixed temperature, $T$, and field--strength, $h_r$, we
performed a disorder average over 1280 samples. An average over such
a large number of samples is necessary because the susceptibility
is highly non self--averaging, as mentioned before. The simulations
were done for fixed ratio $h_r/T$ at different temperatures.

To check equilibration we simulated two replicas of the system:
one starting from an initial configuration with all spins up and
one with all spins down. We assumed that the system has reached
equilibrium when the
magnetization measured for both replicas is the same (within the
errorbars). The time needed for equilibration of all 1280
systems varied much with the system size and temperature --- in
case of the largest size ($L=16$) we used up to $0.5\cdot10^6$ MC--steps for
equilibration and $1.5\cdot10^6$ MC--steps for measurements.
All of the samples were equilibrated for $L<10$. For larger sizes the
number of nonequilibrated samples generally varied between 1\% and 3\%.
The contribution of these samples was estimated to be less than the
error bars in the points so so no significant error was made by
including them. The only exception to this was for $L=16, h_r/ T = 0.5$
for which 5\% of the samples were not equilibrated which gave a
significant error in the susceptibility, though not for the other
quantities. We therefore ignored this data point when analyzing the
susceptibilty.

For each sample and each replica ({\it a,b}) we recorded the
average magnetization per spin $\langle M_{a,b}\rangle$, its square
$\langle M_{a,b}^2\rangle$, the average energy per spin $\langle
E_{a,b}\rangle$
and its square $\langle E_{a,b}^2\rangle$. The angular brackets, $\langle
\ldots \rangle $, denote a thermal average for a single random field
configuration. From these data we get the
specific heat per spin, $C$, the susceptibility $\chi$, the disconnected
susceptibility, $\chi_{{\rm dis}}$, and the order parameter, $m$, as
follows:
\begin{equation}
\begin{array}{clcl}
[C]_{\rm av}=& N\Bigl\{[\langle E^2\rangle]_{\rm av}
-[\langle E\rangle^2]_{{\rm av}}\Bigr\}/T^2\;,\quad
&[m]_{\rm av}&= [\vert\langle M\rangle\vert]_{\rm av}\;, \\
\,&&\quad&\\
\,[\chi]_{\rm av} = &N\Bigl\{[\langle M^2\rangle]_{\rm av}
-[\langle M\rangle^2]_{{\rm av}}\Bigr\}/T\;,\quad
&[\chi_{\rm dis}]_{\rm av}&=N[\langle M\rangle^2]_{{\rm av}}\;,
\end{array}
\end{equation}
where $[\ldots ]_{{\rm av}}$ denotes the average over different random field
configurations.

The procedure we used to extract the critical exponents is the following:
Let $t=T-T_c$ (the deviation from the critical temperature),
then the finite size scaling functions for the above quantities read:
\begin{equation}
\begin{array}{clcl}
\label{sh}
T^2 [C]_{\rm av}&=L^{\alpha/\nu}\tilde{C}(t L^{1/\nu})\;,\quad
&[m]_{\rm av}&=L^{-\beta/\nu}\tilde{m}(t L^{1/\nu})\;,\\
&&\quad&\\
T\,[\chi]_{\rm av}&=L^{2-\eta}\tilde{\chi}(t L^{1/\nu})\;,\quad
&[\chi_{\rm dis}]_{\rm av}&=L^{4-\overline{\eta}}\tilde{\chi}_{{\rm dis}}
(t L^{1/\nu})\;,
\end{array}
\end{equation}
where $\alpha$ is the specific heat exponent, $\beta$ the order parameter
exponent, $\nu$ the correlation length exponent and $\eta$ and
$\overline{\eta}$ describe the power law decay of the connected and
disconnected correlation functions, see e.g.\ \cite{BY}. Note that
the susceptibility exponent is given by $\gamma=(2-\eta)\nu$.
The scaling function
$\tilde{C}(x)$ has a maximum at some value $x=x^*$.
For each lattice--size we estimate
the temperature $T^*(L)$, where $T^2[C]_{\rm av}$ is maximal. Since
$t^*(L)\,L^{1/\nu}=x^*$ we obtain in this way
the critical temperature $T_c$ and the correlation length exponent
$\nu$ from:
\begin{equation}
t^*(L) \equiv T^*(L) - T_c = x^*L^{-1/\nu}\;.\label{tc}
\end{equation}

We denote the value of $T^*(L)^2 C$
at this temperature $T^*(L)$ by $C^*$ and similarly for the
other quantities in Eq. (\ref{sh}).
In the vicinity of $x^*$ the scaling function $\tilde{C}(x)$ can be
approximated by a parabola. Therefore three temperatures near the maximum
of the specific heat are enough to determine the values of $T^*(L)$
as well as $[C^*]_{\rm av}$ etc.
Our results for the exponents obtained in this way are summarized in Table 1.
For illustration, we show the results for $T^*(L)$,
$[C^*]_{\rm av}$, $[\chi^*]_{\rm av}$, $[m^*]_{\rm av}$ and
$[\chi_{\rm dis}^*]_{\rm av}$ for $h_r/T=0.35$
in Fig.\ 1a--d. Several comments have to be made:

1) The higher the field strength the harder it is to equilibrate the samples.
However, the lower the field the less pronounced is the random field behavior
for small lattice sizes because of crossover from pure Ising model behavior.
Therefore the investigation of larger as well as smaller
ratios $h_r/T$ did not seem to be advisable to us. If the transition is
of second order and no tricritical point occurs along the critical
line $(T_c,h_c)$ the exponents should be universal, i.e.\ independent
of the value of $h_r/T$.

2) The shift of $T^*(L)$ with respect to $T_c$ becomes
smaller for low field strength, so it is harder to determine
the exponent $\nu$. In case of $h_r/T=0.25$ it was not possible to
perform an acceptable fit for $T^*(L)$ according to equation
(\ref{tc}).
The values of $\nu$ obtained for the other ratios $h_r/T$ are somewhat
higher than that obtained in \cite{Og}, where $\nu=1.0\pm0.1$.

3) We did not find any indication of a divergence, even logarithmic,
of the specific heat, so $\alpha$ is negative.
This is different from what is found experimentally \cite{BY,Bel}, where
the specific heat diverges logarithmically, corresponding to $\alpha=0$.
Furthermore, in our simulations $\alpha$ seems to get more negative with
increasing ratio $h_r/T$. This may indicate that it is difficult to
determine $\alpha$ when $\alpha$ is negative because non-singular (but
temperature dependent) background terms can give a significant
contribution to the specific heat.

4) The order parameter $[m^*]_{\rm av}$ shows only a very small size
dependence, and does not approach zero but $\lim_{L\to \infty} [m^*]_{\rm av}
\approx 0.52$, $0.50$ and $0.47$
for $h_r/T=0.5$, $0.35$ and $0.25$, respectively (see the inset of fig.\ 1d).
This indicates that $\beta=0$ and that the transition is first order.
This seems to contradict our results for the specific heat since
the specific heat is expected to diverge
as $L^d$ \cite{NN} at a first order transition, because of the latent
heat, whereas our specific heat data seem to
saturate for large $L$. Perhaps the coefficient of $L^d$ is zero (though
we see no symmetry reason for this) or is so small that $L^d$ behavior
would only be seen for larger sizes.

5) For the exponent $\eta$ we get a best estimate that is slightly higher
than $1/2$, which is the value obtained below $T_c$ and also the value {\em at}
$T_c$ if the transition is first order \cite{DSY}. However, the value
$\eta = 0.5$ is not excluded by our data.
For $h_r/T=0.5$ we had to
exclude the size $L=16$ from the analysis since 5\% of our
samples were not  equilibrated and the contribution of these samples was
larger than the error bars.
Our estimates for $\eta$ are consistent
with that obtained in \cite{OgHu}: $\eta=0.5\pm0.1$.

6) The exponent $\overline{\eta}$ for the disconnected susceptibility
turns out to be equal to one, so that the scaling relation
\begin{equation}
\beta=(d-4+\overline{\eta})\nu/2
\label{etabarbeta}
\end{equation}
is fulfilled (as indicated in the table).
We also see that the Schwartz--Soffer \cite{SS} inequality
\begin{equation}
\overline{\eta}\le2\eta\;,
\label{ssineq}
\end{equation}
holds as an equality within the error bars. In \cite{Og} it was found
that $\overline{\eta}=1.1\pm0.1$.

7) In the droplet picture \cite{Vi,Fi} $\theta=2-\overline{\eta}+\eta$
is called the violation of hyperscaling exponent. The hyperscaling
relations then have the spatial dimension, $d$, replaced
by $d-\theta$, e.g.\
\begin{equation}
2-\alpha=(d-\theta)\nu\;.
\label{alphanu}
\end{equation}
As indicated in the table this equality seems {\em not} to be fulfilled
though the error bars are quite large and our estimate for $\alpha$ might
be affected by temperature dependent background terms, as discussed
above.
The estimates of both sides of this equation cannot be
made without knowing {\em both} $\alpha$ and $\nu$ but for $h_r/T=0.25$ we only
determined the ratio. The entries in the table for
$2-\alpha$ and $(d-\theta)\nu$ are therefore left
blank for $h_r /T = 0.25$.

8) One of the main predictions of the droplet picture \cite{Vi,Fi}
is a long tail in the distribution of the susceptibility $\chi$
for samples of size $L$ at $T=T_c$ \cite{DSY}. An analysis of this distribution
extracted from our results for the 1280
samples confirms the existence of this long tail. Figure {\ref{fig2}}
shows the histograms for the probability distribution $P(\chi)$ close to the
temperature $T^*$ (T=3.80 for L=8 and T=3.75 for
L=16) for $h_r/T$=0.35. The second moment of this distribution
$[{\chi^*}^2]_{\rm av}$,
shown in the inset of fig.\ 1c, scales like $L^\zeta$,
with $\zeta=3.8\pm0.1$ (for $h_r/T=0.35$),
which is larger
than the square of the mean $L^{4-2\eta}\sim L^{2.9}$, but somewhat
smaller than the predicted value $\zeta=6-\overline{\eta}-\eta\approx4.4$
\cite{DSY}.
We attribute this difference in $\zeta$ to the
number of samples being too small to catch a sufficient number of rare
samples which dominate the higher moments.

To conclude, while the data for the magnetization and
disconnected susceptibility indicate a first order transition
fairly convincingly, the specific heat seems to saturate to
a finite value so there is no detectable latent heat.
It is interesting to ask if the order of the
transition might be different for a different random field distribution,
since mean field theory predicts \cite{a78} that
the transition becomes first order for large fields
for the $\pm h$ distribution, but not, for example, for the Gaussian
distribution.  Since the multi-spin coding
technique that we used does not work for a continuous distribution of
fields, the answer to this question needs an even larger computing
effort.  Nevertheless we are currently attempting to carry out similar
calculations for the Gaussian distribution.
Our results are
consistent with the Schwartz-Soffer inequality, Eq. (\ref{ssineq}),
being satisfied as
an equality, and support the scaling relation, Eq. (\ref{etabarbeta}). The
scaling relation involving the specific heat, Eq. (\ref{alphanu}), does not
seem to be satisfied, though our values for $\alpha$ may only be
effective exponents particularly since we find $\alpha$ is negative and
so a more detailed determination of non-singular background terms
might be necessary to determine
$\alpha$ accurately. Our results do support the prediction of the droplet
theory that there are large sample to sample variations in the
susceptibility at $T_c$.

We would like to thank D. P. Belanger for helpful discussions.
One of the authors (HR) would like to thank the HLRZ at the KFA J\"ulich
(Germany) for allocation of computer time for the $L=24$ run and
acknowledges financial support from the DFG (Deutsche Forschungsgemeinschaft).
The work of APY was supported in part by the NSF grant no.\ DMR-91-11576.

\newpage

\figure{\label{fig1}\baselineskip=18pt
The results of least square fits to data obtained by the procedure
described in the text for $h_r/T=0.35$.
The points indicated by diamonds ($\diamond$) correspond to lattice sizes
L=$4$, $6$, $8$, $10$, $12$, $16$ --- from right to left in (a) and (b) and
left to right in (c) and (d). With the exception of the susceptibility we also
inserted data for $L=24$ with squares ($\Box$), which we obtained by using a
the same algorithm but on a CRAY Y-MP instead of the transputer array and
which are averaged over only 64 samples. The $L=24$ data were not used
for the least square fits.
(a) The temperature $T^*(L)$ of the specific heat maximum
versus $L^{-1/\nu}$ with $\nu=1.64$ and $T_c=3.552$, see Eq. (\ref{tc}).
(b) Specific heat $[C^*]_{\rm av} \equiv L^{\alpha/\nu}  \tilde{C}(x^*)$
versus $L^{-1/\nu}$ with $\alpha=1.04$
and $\nu$ as in (a), see Eqs. (\ref{sh}) and (\ref{tc}).
The specific heat appears to saturate for $L \to
\infty$ at a value of $\lim_{L\to \infty} [C^*]_{\rm av} \simeq 25.3$.
(c) Susceptibility $[\chi^*]_{\rm av}  \equiv L^{2-\eta} \tilde{\chi}(x^*)$
in a log--log plot. The slope
of the straight line is $2 - \eta$ with $\eta = 0.53$.
The inset shows the second moment
$[{\chi^*}^2]_{\rm av}$ of the probability distribution $P(\chi)$
in a log--log plot. The slope of the straight line is $\zeta=3.82$.
(d) Disconnected susceptibility $[\chi_{\rm dis}^*]_{\rm av}
\equiv  L^{4-\overline{\eta}}\tilde{\chi}_{dis}(x^*)$
in a log--log plot. The slope is $4 - \overline{\eta}$ with
$\overline{\eta} = 1.0$. The insert shows the magnetization
$[m^*]_{{\rm av}} \equiv L^{-\beta/\nu} \tilde{m}(x^*)$
as a function of a $L$ (the scale of the x--axis
is logarithmic). The straight line is the extrapolation to
$[m^*]_{{\rm av}}(L=\infty)$, which is clearly nonzero and so $\beta =
0$.}

\figure{\label{fig2}\baselineskip=18pt
The histograms for the probability distribution $P(\chi)$ of the
susceptibility for different $L = 8$ and 16 with $h_r/T=0.35$.
The temperatures are chosen to be as close to $T^*(L)$ as possible:
$T=3.80$ for L=8 and $T=3.75$ for L=16.
\\
The y--axes of the inserts are scaled differently to emphasize
the long tail of the distribution. This feature originates
in the rare samples with the extremely large
values of the susceptibility scaling with the volume of the system
(since $4-\overline{\eta}\approx3$).}

\mediumtext
\begin{table}
\caption{The critical exponents obtained via finite size scaling
according to the procedure described in the text.}
\begin{tabular}{|c||rl|rl|rl|}
$h_r/T$    & $\quad$ 0.25 && $\quad\qquad$ 0.35 && $\quad\qquad$ 0.5 & \\
\hline\hline
$T_c$             & 3.9$\;\;$&$\pm$0.1   & 3.55&$\pm$0.05 & 3.05&$\pm$0.05 \\
\begin{tabular}{c}
$\;\;\nu$\\
$\;\;\alpha$
\end{tabular}
&
$\bigg\}\frac{\textstyle\alpha}{\textstyle\nu}$=--0.50&$\pm$0.05
&
\begin{tabular}{r}
1.6\\
--1.0
\end{tabular}
&
\begin{tabular}{l}
$\pm$0.3\\
$\pm$0.3
\end{tabular}
&
\begin{tabular}{r}
1.4\\
--1.5
\end{tabular}
&
\begin{tabular}{l}
$\pm$0.2\\
$\pm$0.3
\end{tabular}
\\
$\eta$            & 0.60&$\pm$0.03 & 0.56&$\pm$0.03 & 0.6$\;\;$&$\pm$0.1 \\
$\overline{\eta}$ &0.97&$\pm$0.08 & 1.00&$\pm$0.06 & 1.04&$\pm$0.08 \\
$\theta=2-\overline{\eta}+\eta$
                  & 1.6$\;\;$&$\pm$0.1  & 1.6$\;\;$&$\pm$0.1 &
1.6$\;\;$&$\pm$0.1 \\ \hline
$\beta$           & 0$\quad$& & 0$\quad$& & 0$\quad$& \\
$(d-4+\overline{\eta})\nu/2$
                  & 0$\quad$&$\pm$0.05 & 0$\quad$&$\pm$0.05 &
0$\quad$&$\pm$0.05 \\ \hline
$(d-\theta)\nu$   & &  & 2.3$\;\;$&$\pm$0.5 & 2.0$\;\;$&$\pm$0.6 \\
$2-\alpha$        & & & 3.0$\;\;$&$\pm$0.3 & 3.5$\;\;$&$\pm$0.3 \\
\end{tabular}
\end{table}
\vfill
\eject
%
%***************************************************************************
%*                                                                         *
%*  Figure 1a                                                               *
%*                                                                         *
%***************************************************************************
%
% GNUPLOT: LaTeX picture
\setlength{\unitlength}{0.240900pt}
\ifx\plotpoint\undefined\newsavebox{\plotpoint}\fi
\sbox{\plotpoint}{\rule[-0.175pt]{0.350pt}{0.350pt}}%

\begin{center}
\Large Fig.\ 1d
\end{center}
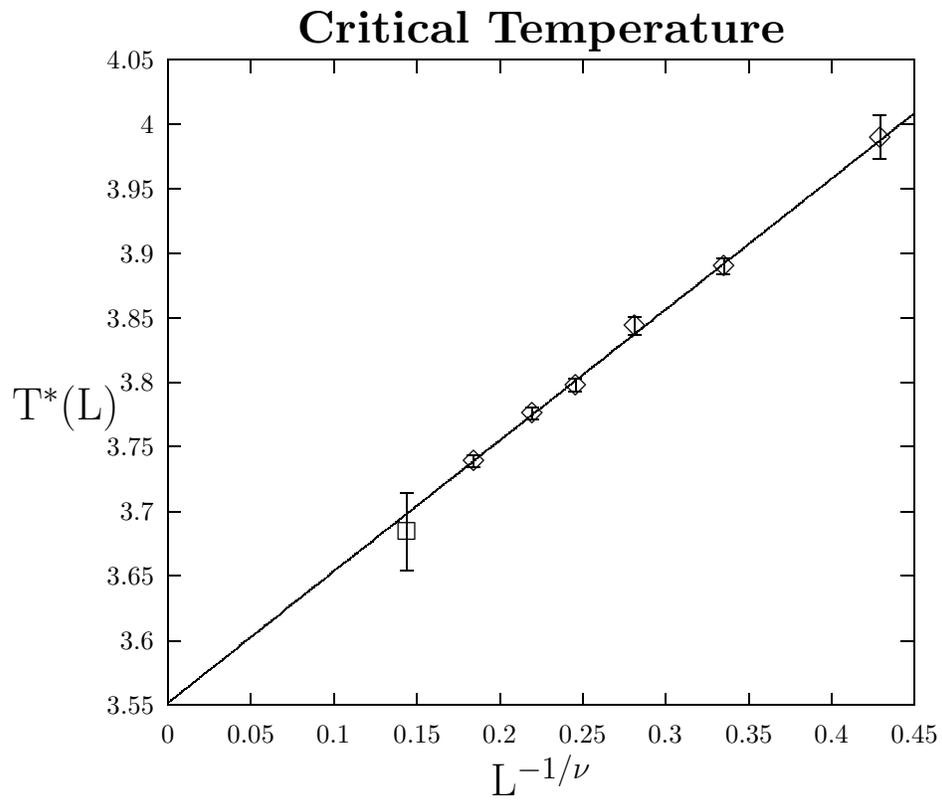
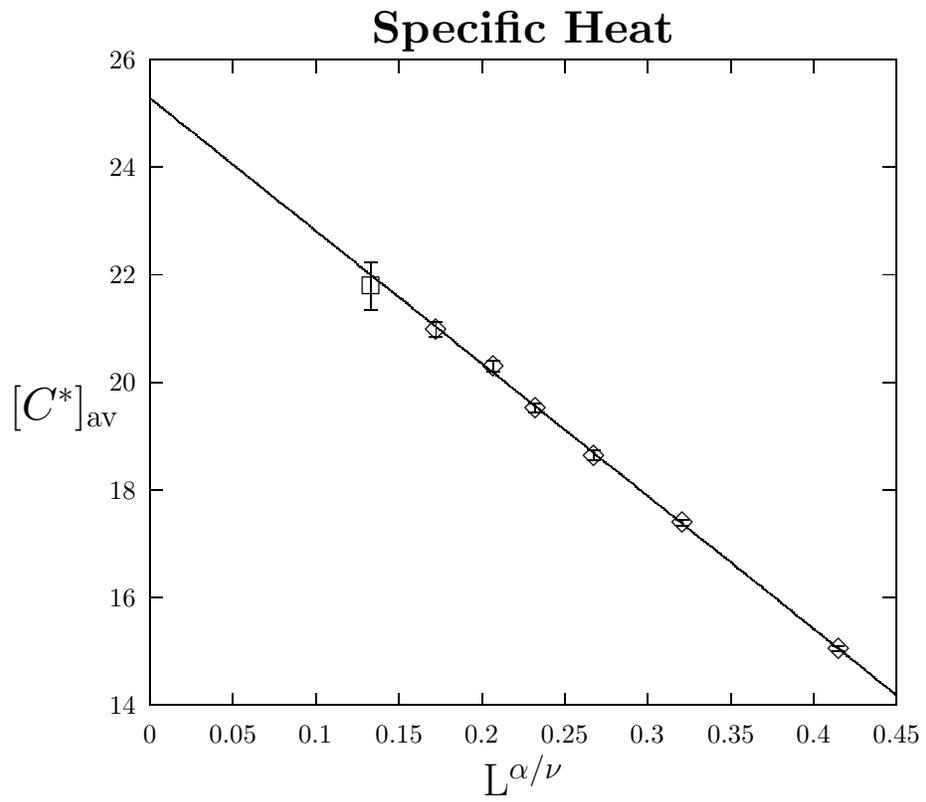
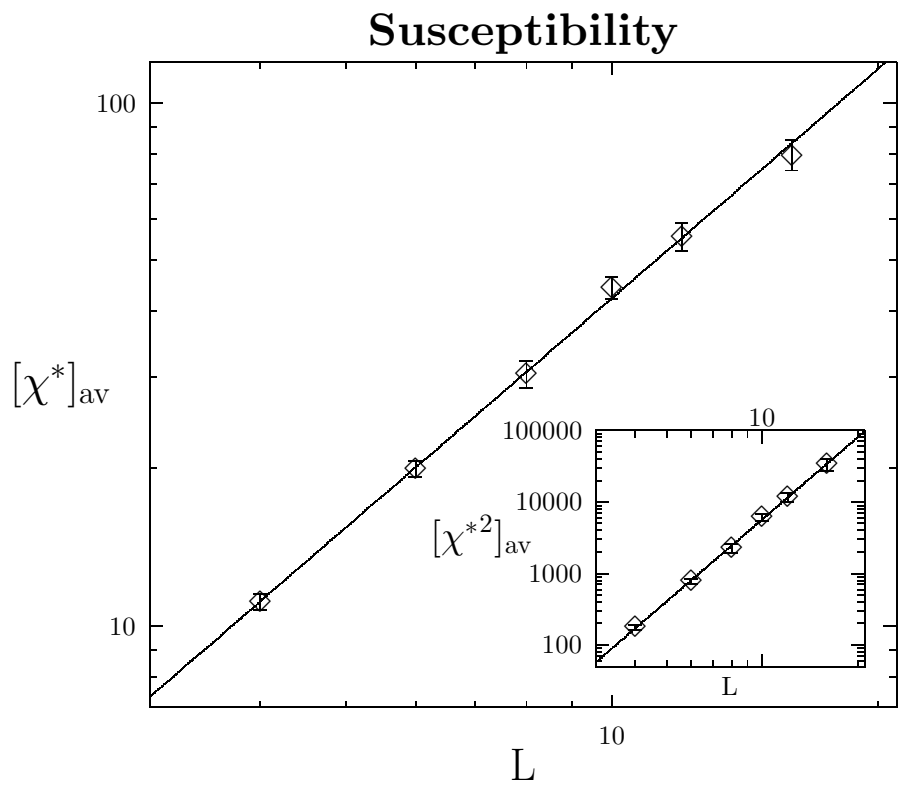
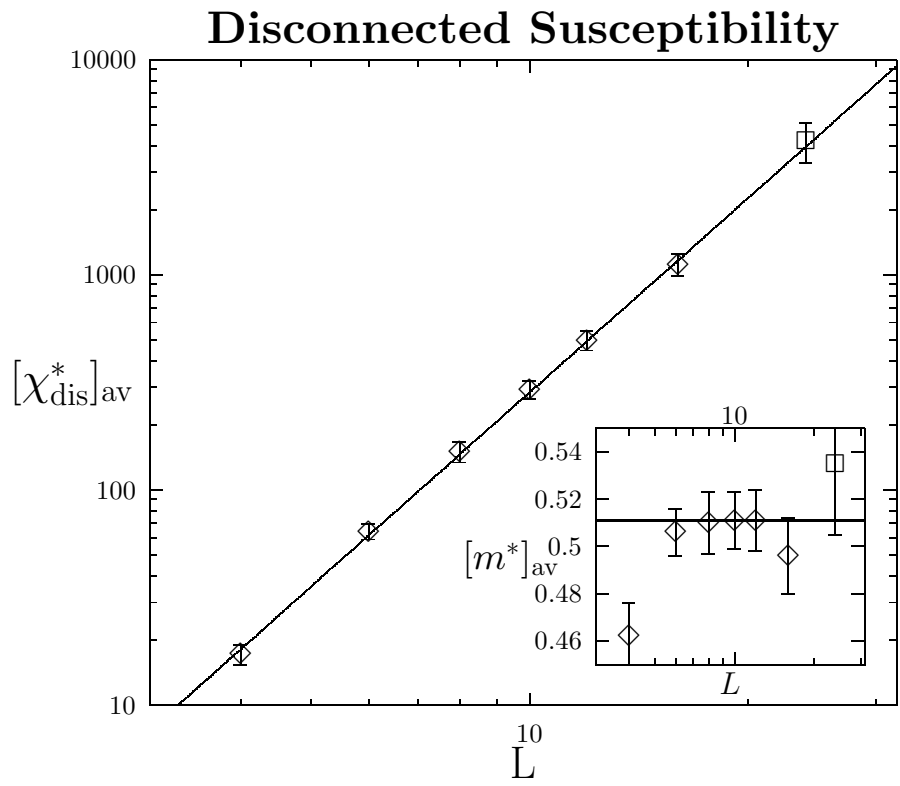
\end{document}